\documentclass[12pt]{iopart} 
\usepackage{epsf}
\usepackage{epsfig} 
\usepackage{amssymb,amsthm}
\usepackage{iopams}
\def\eqref#1{(\ref{#1})} 
\newcommand{\abs}[1]{|#1|}
\newcommand{\calC}{{\cal{C}}} 
\newcommand{\bbR}{{\mathbb{R}}} 
\begin{document} 
\title[Delay Equations]
{Delay Equations and Radiation Damping}
\author{C.~Chicone\dag, S.~M.~Kopeikin\ddag, B.~Mashhoon\ddag\   
and D.~G.~Retzloff\S}
\address{\dag\ Department of Mathematics,
University of Missouri,\\ Columbia, MO 65211, USA;
chiconec@missouri.edu} 
\address{\ddag\ Department of Physics and Astronomy, 
University of Missouri,\\ Columbia, MO 65211, USA;
kopeikins@missouri.edu; mashhoonb@missouri.edu} 
\address{\S\ Department of Chemical Engineering, 
University of Missouri,\\ Columbia, MO 65211, USA;
retzloffd@missouri.edu} 
\begin{abstract} 
Starting from delay equations that model field retardation effects,
we study the origin of runaway modes that appear in the solutions 
of the classical
equations of motion involving the radiation reaction force.
When retardation effects are small, we argue that
the physically significant solutions belong to the so-called
\emph{slow manifold} of the system and we identify this invariant manifold
with the attractor in the state space of the delay equation.
We demonstrate via an example that when retardation
effects are no longer small, the motion could exhibit \emph{bifurcation}
phenomena that are not contained in the local equations of motion. 
\end{abstract} 
\pacs{0430,0350,0230K}
\maketitle
\section{Introduction}\label{sec:1} 
In the treatment of the motion of \emph{extended} bodies in classical 
field theory, the 
derivation of radiation reaction forces 
is based upon certain expansions
of the retarded field potentials in powers of the retardation~\cite{lorentz}.
The resulting local equations of motion involve derivatives of the acceleration
and generally suffer from the existence of 
unphysical runaway solutions. 
Under certain model circumstances, we trace the origin of these problems to
the expansion of the functions of the retarded 
arguments resulting in the replacement
of the original nonlocal delay equations of motion by
local higher-derivative equations that exhibit runaway solutions.
In this general context, the properties
of the delay equations that appear in classical field theory
were first studied
by L.~Bel~\cite{bel,bel2,bel3}.
Although our approach is rather general, for the sake of concreteness
we discuss physical situations involving only the gravitational
interaction.  

Consider, for instance, inspiraling 
compact binaries that are expected to be promising
sources of gravitational radiation. 
For a binary that is comprised of two 
compact objects---neutron stars or black holes---with, say,
approximately equal 
masses $m$ and $m'$ in nearly circular orbits about each other, 
the relative orbital radius decays because orbital energy is emitted in the
form of gravitational radiation. The dynamics of a usual
binary system can be adequately described using the
post-Newtonian approximation scheme
that is valid in case the gravitational field is everywhere `weak' 
and the motion is slow, that is $v\ll c$, where $v$ is the
characteristic orbital speed and $c$ is the speed of light. 
Although Einstein's equations have a hyperbolic character associated
with the retarded gravitational interaction,
the standard post-Newtonian approximation scheme of general
relativity deals with
functions of instantaneous coordinate time $t$ rather than the
retarded time $t_r=t-r/c$, where (for the binary
system) $r$ is the effective distance between the bodies (approximately
the relative orbital radius).
The gravitational potentials, which are originally functions of
the retarded time, are expanded in Taylor series about $t$ using
the effective small parameter in this expansion that 
can be written as $\omega_b\, r/c=v/c$, 
where $\omega_b=2\pi/P_b$ is the relative
orbital frequency and $P_b$ is the binary period. 
Because the gravitational waves emitted by the
binary have an effective frequency of $\approx 2\omega_b$
and wavelength $\lambda_b\approx c P_b/2$, the small parameter
in the expansion can be reduced to the ratio $\pi r/\lambda_b$.
Due to the observational fact that in
typical astronomical systems $v/c\ll 1$, the
first few terms of such an expansion can be used
to derive the post-Newtonian equations of motion that
describe, for instance, the 
orbital evolution of the binary pulsars 
discovered by Hulse and Taylor~\cite{ht}. 
The post-Newtonian 
equations of motion of binary stars are similar to the Abraham-Lorentz 
form in electrodynamics~\cite{barut,barut2,barut3}
but, because of the tensorial character of the gravitational field,
these equations involve not only the third, but the fourth and fifth 
derivatives of the stars' positions with respect to time as well. 
Schematically, the equation of the relative orbital motion reads
\begin{equation}
\label{5der}
\fl\ddot{\bf r}={\bf F}_0({\bf r})
+c^{-2}{\bf F}_2({\bf r},\dot{\bf r},\ddot{\bf r})
+c^{-4}{\bf F}_4({\bf r},\dot{\bf r},\ddot{\bf r},{\bf r}^{(3)},{\bf r}^{(4)})
+c^{-5}{\bf F}_5({\bf r},\dot{\bf r},\ddot{\bf r},{\bf r}^{(3)},
{\bf r}^{(4)},{\bf r}^{(5)}), 
\end{equation}
where ${\bf r}$ is the radius vector connecting the stars,
the overdot denotes differentiation with respect to time, 
${\bf r}^{(n)}:= d^n{\bf r}/dt^n$  and per reduced mass 
${\bf F}_0({\bf r})=-(G(m+m')/r^3){\bf r}$ is the Newtonian force,  
${\bf F}_2$ is the post-Newtonian force, 
${\bf F}_4$ is the post-post-Newtonian force 
and ${\bf F}_5$ is the gravitational radiation reaction force
responsible for the decay of the orbital period ($\dot{P}_b<0$)
associated with the emission of gravitational waves by the 
binary system. In the quadrupole approximation under consideration
here, the gravitational waves carry away energy and angular
momentum, but not linear 
momentum~\cite{damour,damour2, schafer, sergei,
leon, kostas, cmr,cmr2,cmr3,cmr4,cmr5};
therefore, the total momentum of the binary system is
conserved and this fact is responsible for the absence of a force
term proportional to $c^{-3}$ in~\eqref{5der}. Moreover,
all tidal, spin-orbit, and spin-spin interactions are
neglected in~\eqref{5der}; the only
parameters in equation~\eqref{5der} are the masses
and the separation between the centers of mass 
of the members of the binary system. 
We mention that relativistic hydrodynamical (Euler) equations similar
to system~\eqref{5der} have been derived to
describe the motion of the fluid elements of the 
stars~\cite{rezzolla, rezzolla2}.

The higher time-derivative equations of the form~\eqref{5der}
cannot be used directly to predict    
the dynamical evolution of a physical system because of the
existence of so-called runaway modes that have been much 
discussed in the literature on electrodynamics but not in
connection with astrophysical 
problems involving gravitational radiation reaction and 
the calculation of templates of the gravitational waves 
emitted by coalescing 
binaries~\cite{rezzolla, rezzolla2, rezzolla3, rezzolla4}. 
In analogy with electrodynamics,
the existence of these runaway modes suggests that 
the truncated equations of the form~\eqref{5der} may not
correctly predict the qualitative behavior of the solutions of the 
original true dynamical delay-type equations~\cite{bel} 
that involve the retarded time $t_r=t-r/c$.
Moreover, the existence of runaway modes can cause
serious difficulties for numerical integration in addition
to the problems associated with the 
inaccuracies inherent in the approximation of
higher-order derivatives by finite differences~\cite{rezzolla,rezzolla2}.

In case the post-Newtonian expansion parameter 
$r/\lambda_b$ is sufficiently small, we will provide in
the following section a 
theoretical basis for eliminating the runaway solutions by replacing
system~\eqref{5der} with a new model that is a system of 
second-order ordinary differential equations. 
Within this theory, high-order vector 
differential equations like equation~\eqref{5der} are not the desired 
approximate equations of motion, and they should
not be used for numerical integration. 
Rather, system~\eqref{5der} must be viewed merely as an intermediate
step in the derivation of the physically correct, second-order
model equation with no runaway solutions 
that faithfully approximates the dynamics of the underlying 
delay-type equation. 
For illustration purposes, we apply this 
approach in section~\ref{sec:3} to a discussion of one-dimensional
gravitational dynamics of a two-body system.
As expected, the reduced model predicts the correct
motion of the binary except possibly when
$r/\lambda_b$ is not small and residual terms that have
been neglected in equation~\eqref{5der} start to
play a significant role. 
Indeed,
for an inspiraling binary system, the effective delay
$\omega_br/c$ increases to some 
noticeable finite value as the system approaches
coalescence.
Motivated by this physical scenario, we introduce a simple model
involving variable delay in section~\ref{sec:4} that can be expressed
as a Duffing-type differential-delay
equation. This model is then analyzed 
to show some specific behavior of this delay equation that
is not predicted by expansion in powers of the delay.
Finally, section~\ref{sec:5} contains a discussion
of our results.

\section{Delay equations with small delays}\label{sec:2} 
The delay-type equations of motion 
with retarded arguments are usually too complicated 
for mathematical analysis; therefore, we limit our discussion in
this section to equations with constant delays. Although this
is an unrealistic restriction in general,
we note that for an astrophysical binary
system consisting of compact point-like neutron
stars or black holes moving around each other along
circular orbits, the delay is almost constant.
In fact, a close approximation to this delay
is the ratio $r/c$, where $r$, the
radius of the relative orbit, is changing
very slowly due to the emission of gravitational
energy in the form of gravitational waves.

Taking into account the last remark,
let us consider a family of delay differential equations of the form
\begin{equation}\label{RFDE}
\dot x(t)=F(x(t-\tau),x(t)),
\end{equation}
where $\tau$ is viewed as a real dimensionless 
parameter and $x$ is a variable
in $\bbR^n$; intuitively, the constant delay $\tau$ corresponds in
effect to $\omega_b\,r/c$. 
The members of this family are examples of a more general and widely
studied class called retarded functional differential equations 
(see~\cite{diek, hale}). 

Using the delay equation~\eqref{RFDE} as an abstraction for the
retarded-time model that is supposed to be approximated by a system
of the form~\eqref{5der},  we will discuss an approach for extracting the 
`correct' dynamical equations of motion from system~\eqref{5der} that
eliminates the runaway solutions. 

Our approach assumes the existence of an attractor for 
the underlying delay-type equation.
We will rely on the work of Bel~\cite{bel} for (numerical) evidence in favor of
the existence of attractors in the retarded equations of motion 
with space-dependent delays that appear in electrodynamics; but, 
we know of no mathematical proof for the existence of attractors for these
equations or for the similar delay-type equations of astrophysics. 
Indeed, the proof of the existence of attractors for
delay-type equations with space-dependent delays 
remains a challenging mathematical problem of physical significance.
For the delay equation~\eqref{RFDE}, however, 
if $\abs{\tau}$ is sufficiently small, 
then the corresponding member of the family~\eqref{RFDE}
has a global $n$-dimensional attractor such that the restriction of
the delay equation to this attractor is equivalent to 
a first-order system ${\cal S}_A$
of ordinary differential equations.
We will eventually outline a proof of this result.
But, let us first discuss our approach
to eliminating the runaway solutions.

The solutions of the delay equation~\eqref{RFDE} 
approach the attractor exponentially fast; therefore, the system
${\cal S}_A$ on this attractor 
determines (asymptotically) the true dynamical behavior of the system, 
hence we consider it to be the `correct' physical model. 
On the other hand, it is easy to see that
if equation~\eqref{RFDE} 
is expanded in the small parameter $\tau$ and truncated at some order $N$, then
an $N$th-order ordinary differential system ${\cal S}_N$  
akin to system~\eqref{5der}
is obtained such that the coefficient of the 
$N$th-order time derivative of $x$ contains the factor $\tau^N$ and is
therefore singular in the limit as $\tau\to 0$.

For $\tau>0$ and sufficiently small,  the high-order differential
equation ${\cal S}_N$ has an equivalent
first-order system ${\cal S}$ 
that has an $n$-dimensional (invariant) slow manifold. 
Moreover, this slow manifold has corresponding
stable and unstable manifolds; in effect, 
the first-order system has (physical) solutions
that are asymptotically attracted to the slow manifold and
(unphysical or runaway) 
solutions that are asymptotically repelled from the slow manifold. 
The restriction of the first-order 
singularly perturbed system of differential equations
to its slow manifold is of course an $n$-dimensional
first-order system
of ordinary differential
equations ${\cal S}_S$ on this $n$-dimensional manifold. 
Our main result states that 
\emph{in appropriate local coordinates, the system ${\cal S}_S$ on
the slow manifold agrees to order $N$ in $\tau$ 
with the first-order system ${\cal S}_A$
on the global attractor of the underlying 
delay differential equation; therefore, the system ${\cal S}_S$,
which can be obtained directly from the high-order differential
equation ${\cal S}_N$,
is a faithful approximation of the `correct' physical model.} 
Generalizing to the Abraham-Lorentz type equation~\eqref{5der}
(analogous to ${\cal S}_N$), the 
correct physical model is obtained as the system of ordinary
differential equations (analogous to ${\cal S}_S$) 
that determines the motion on the slow
manifold of a corresponding first-order system
that is viewed as being singularly perturbed relative to the 
small parameter $r/\lambda_b$.

While there is evidence that the mathematical assertions in the scenario 
just proposed are valid, some of these assertions have not yet been
rigorously justified in full generality, even for the case of fixed delays.
In the remainder of this section we will provide some evidence,
in the case of fixed delays, for the
existence of a global attractor and for the claim that the dynamical
system on this attractor is well approximated by the dynamical
systems on the slow manifolds of singularly perturbed
first-order systems obtained by truncations of the expansion of
the delay equation in powers of the delay.  

Our approach for the elimination of runaway solutions 
is equivalent to the procedure 
of iterative reduction (also called order reduction)
that is often used to 
eliminate runaway solutions by means of the evaluation of the 
higher time-derivative terms
in equations like (\ref{5der}) 
by the repeated substitution of the equations of
motion and the subsequent reduction of the resulting 
equation to one of the second order  (cf.~\cite{barut,barut2, barut3}).
Thus, our approach provides
a theoretical framework for the rigorous justification of iterative
reduction (cf.~\cite{barut, barut2, barut3}), a procedure that has
been justified so far by physical intuition.
    
For \emph{conservative} higher time-derivative systems,
the order reduction procedure has been investigated within
the frameworks of Lagrangian and Hamiltonian dynamics by
a number of authors (see~\cite{spohn, eli, sim,dam,spohn2} 
and the references cited therein).
In particular, it can be shown that under suitable conditions---relevant, 
for instance, to Euler-Lagrange equations analogous to equation~\eqref{5der}
truncated at the fourth order---higher-derivative Lagrangians can
be iteratively reduced by  redefinitions of position variables~\cite{dam}.

Returning to the delay equation~\eqref{RFDE}, we note that
it has an infinite-dimensional state space of initial
conditions.
For example, if the delay $\tau$ is a fixed positive number, then the 
natural state space of initial conditions 
is the infinite-dimensional vector space of continuous
$\bbR^n$-valued functions on the interval $[-\tau,0]$. This
space endowed with the supremum norm is a Banach space that we will
denote by $\calC$. 
Note that for an arbitrary  continuous $\bbR^n$-valued
function $\gamma$ defined on the interval $[t-\tau,t]$, the function 
$\gamma_t$ given by $\gamma_t(\vartheta)=\gamma(t+\vartheta)$ is in $\calC$. 
Under the
assumption that $F$ is a smooth function and 
$\phi\in\calC$,
there is a unique continuous solution $y$ of the corresponding
delay equation in the family~\eqref{RFDE} such that $y$ is 
uniquely defined for $t\ge -\tau$ and $y_0=\phi$ 
(see, for example, \cite{diek, hale}).
The state of the system at time $t>0$ is defined to be the function $y_t$ in
$\calC$.

To see that there is an attractor for the 
family~\eqref{RFDE} in case $\tau$ is sufficiently small, it is
convenient to 
introduce the fast time $s:=t/\tau$,
valid for $\tau\ne 0$,
so that with $y(s):=x(t)$ the family~\eqref{RFDE} takes the form
\begin{equation}\label{fRFDE}
y'(s)=\tau F(y(s-1),y(s))
\end{equation}
and each member of this family, parametrized by $\tau$,
has the same state space---the
continuous functions on the interval $[-1,0]$. For each
$\tau\ne 0$ the delay equation~\eqref{fRFDE}
is equivalent to the corresponding member of the family~\eqref{RFDE}.
For $\tau=0$ the corresponding differential equations are not
equivalent, but this is of no consequence because we are 
only interested in the solutions of the
family~\eqref{RFDE} for $\tau\ne 0$.
By viewing the unperturbed system~\eqref{fRFDE}, namely  $y'(s)=0$,
as a delay equation with unit delay, it is clear that the solution with 
initial state $\phi$ is given by $y=\phi$ on the interval $[-1,0]$
and by the constant $y=\phi(0)$ for $t \ge 0$. The initial
state in $\calC$ thus evolves at time $t=1$ to
its final constant state, the function defined on the interval $[-1,0]$ 
with the constant value $\phi(0)$. 
Thus, we conclude that the
$n$-dimensional space of constant functions on $[-1,0]$ is a global attractor
for the delay equation $y'(s)=0$.
Moreover, the convergence to this attractor is faster than any exponential
(the solution reaches the attractor in finite time),
and the dynamical system on this attractor is given by the ordinary
differential equation $y'(s)=0$. If $F$ is appropriately bounded
and $\tau$ is sufficiently small, then, because the contraction rate to
the attractor is exponentially fast,
the attractor persists in the family~\eqref{fRFDE} 
in analogy with the persistence of attractors in 
finite-dimensional dynamical systems; in fact,
each corresponding member of the family~\eqref{fRFDE} has an
$n$-dimensional attractor in the state space $\calC$ and the restriction
of the dynamical system to this attractor is an ordinary differential equation.
In particular, the family~\eqref{fRFDE} has a corresponding family
of invariant manifolds (that is, manifolds consisting of a
union of solutions) that depend smoothly on the parameter $\tau$.

To identify the dynamical system on an attractor of a delay equation,
let us suppose that the delay equation~\eqref{RFDE} has
a family of $n$-dimensional invariant manifolds parametrized 
by $\tau$. Moreover, let $\xi$ denote the local coordinate on
these invariant manifolds, and let $x(t,\xi,\tau)$  denote the solution
with the initial condition $x(0,\xi,\tau)=\xi$ 
on the invariant manifold corresponding to the parameter value $\tau$.
Because these solutions satisfy the delay equation~\eqref{RFDE}, we have
that 
\begin{equation}\label{sfeq}
\dot x(t,\xi,\tau)=F(x(t-\tau,\xi,\tau), x(t,\xi,\tau));
\end{equation} 
therefore, the generator of the dynamical system on the attractor is the
vector field 
\begin{equation}\label{infgen}
X(\xi,\tau):=\frac{\partial}{\partial t} x(t,\xi,\tau)\big |_{t=0}
=F(x(-\tau,\xi,\tau),\xi).
\end{equation}
Under our assumption that this vector field is analytic in $\tau$, 
it can be expanded 
as a Taylor series about $\tau=0$ by differentiating the
function $F(x(-\tau,\xi,\tau),\xi)$ with respect to $\tau$.
To this end, we note that the partial derivatives
of $x(-\tau,\xi,\tau)$ with respect to its first
argument can be evaluated using equation~\eqref{sfeq}, and
partial derivatives with respect to its third argument vanish
at $\tau=0$ since $x(0,\xi,\tau)=\xi$; moreover,
its mixed partial derivatives can be evaluated by 
differentiation of equation~\eqref{sfeq} with respect to $\tau$. 

As a concrete and instructive example of the construction
of the dynamical system on an attractor, let us
consider a simple case of equation~\eqref{RFDE} by replacing it with
$\dot x(t)=\hat f(x(t-\tau))$ where the 
function $\hat f:\bbR\to \bbR$ is the scalar linear function
given by $\hat f(x)=a x$ so that the associated
family of delay equations is
\begin{equation}\label{firstfam}
\dot x(t)=a x(t-\tau).
\end{equation}
In this case, there is a corresponding family of solutions given by
\begin{equation}\label{eq7} 
x(t,\xi,\tau)=e^{\lambda(\tau) t} \xi, \end{equation}
where $\lambda(\tau)$ is the unique \emph{real} root
of the equation $\lambda=a \exp(-\lambda \tau)$,
a fact that is easily checked by direct substitution of
equation~\eqref{eq7} into the delay equation~\eqref{firstfam}.

The dynamical system on the invariant manifold 
is generated by the family of vector fields
$
X(\xi,\tau)=\hat f(x(-\tau,\xi,\tau))
=ax(-\tau,\xi,\tau)=a \exp(-\lambda(\tau) \tau)\xi
 =\lambda(\tau) \xi
$.
By an application of the Lagrange inversion formula~\cite{aar},
the Taylor series expansion of $X(\xi,\tau)$ about $\tau=0$ is 
\begin{equation}\label{eq:ffe}
X(\xi,\tau)= \xi\sum_{n=1}^\infty \frac{(-1)^{n-1} n^{n-1} a^n}{n!} \tau^{n-1},
\end{equation}
and its radius of convergence is $\tau^*= (|a| e)^{-1}$.
The qualitative behavior of solutions of the system~\eqref{firstfam}
for small $\tau$ is clear: all solutions are attracted
to a one-dimensional invariant manifold on which the dynamical
system is the linear ordinary differential equation
$\dot x=X(x,\tau)$. For example,
if $a<0$ and $\tau$ is sufficiently small,
then all solutions are attracted to the trivial solution $x(t)\equiv 0$.

Let us now turn to the standard approach in physics where
an underlying delay equation is expanded in powers of the
delay to obtain an ordinary differential equation of motion.
To illustrate this, let us consider a special scalar case of
the delay equation~\eqref{RFDE} given by
\begin{equation}\label{fgfam}
\dot x=f(x(t-\tau))+g(x(t)),
\end{equation}
and let us suppose, in analogy with the true dynamical delay-type equations
that might arise in theories of electromagnetism and gravitation,
that the true equation of motion for some
process is the delay equation~\eqref{fgfam}. 
The result of expanding equation~\eqref{fgfam}
to order $\tau^2$ is the second-order differential equation
(an analogue of equation~\eqref{5der})
\begin{equation}\label{trun1}
\dot x=f(x)+g(x)-\tau f'(x)\dot x
+\frac{\tau^2}{2} [f''(x)\dot x^2+f'(x)\ddot x],
\end{equation}
where a prime denotes differentiation with respect to $x$.
Although we only write the second-order expansion, we note
that the coefficient of the $N$th-order time derivative of
$x$ in the $N$th-order expansion is $(\tau^N/N!)f'(x)$. Hence, the
corresponding $N$th-order ordinary differential equation
is singular in the limit as $\tau\to 0$. 
Also, if we assume that system~\eqref{fgfam} has a 
(smooth) family of attractors parametrized by
$\tau$, then the corresponding family of vector fields generating
the dynamical systems on these attractors is given to second order
in $\tau$ by
\begin{eqnarray}\label{eq:vectfield}
\nonumber X(x,\tau)&=&f(x)+g(x)-\tau f'(x)(f(x)+g(x))\\
\nonumber &&\mbox{}
+\frac{1}{2} \tau^2 \big\{f''(x)[f(x)+g(x)]^2\\
&&\mbox{}+f'(x)[3f'(x)+g'(x)][f(x)+g(x)]\big\}+O(\tau^3).
\end{eqnarray}

The `correct' model (that is, the dynamical system on
the attractor in the original delay equation) 
can be obtained by treating the expanded and truncated system akin
to system~\eqref{trun1} as a singular perturbation problem, 
which can be analyzed using 
Fenichel's geometric theory of singular perturbations~\cite{fen, fen2}.
A basic result of this theory states that if an $N$th-order
singular perturbation problem with small parameter $\tau$ is recast as a 
first-order (`fast') system and the corresponding   
unperturbed system has an invariant manifold that 
satisfies certain conditions (normal hyperbolicity), 
then for sufficiently small $\tau$
each member of the family of perturbed first-order systems has an 
invariant slow manifold. The dynamical
system on this slow manifold for the perturbed first-order
family obtained from the $N$th-order truncation of the delay equation
is the desired faithful approximation to the correct model.
For example, let us recast the second-order 
ordinary differential equation~\eqref{trun1} as the first-order 
singular perturbation problem
\begin{eqnarray}\label{vecsingsys}
\nonumber\dot x&=& u,\\
\tau^2 \dot u &=& (f'(x))^{-1} 
\big[2(1+\tau f'(x))u-2 f(x)-2 g(x) 
-\tau^2 f''(x)u^2\big]. 
\end{eqnarray}
Using Fenichel's theory, it is easy to show that 
each member of this family, corresponding to 
a sufficiently small value of $\abs{\tau}$, has a slow manifold.
Also, it is possible to prove that the family of vector fields on these
manifolds agrees to order $\tau^2$ with 
the family~\eqref{eq:vectfield} of vector fields on the attractor 
in the state space of the underlying family of delay equations~\eqref{fgfam}.
We note that these results are valid for the vector case
of delay equation~\eqref{fgfam} as well.

For the delay equation~\eqref{fgfam}, 
and also for more general families of delay equations 
where the delay is viewed as a small parameter, 
we conjecture that the slow vector field, 
for an  appropriately defined first-order system
that is equivalent to the $N$th-order truncation of the expansion of the family 
in powers of the small delay, agrees to order $N$ with the vector field
on the attractor in the state space of the original delay equation.
We have just mentioned that this conjecture is true for
the delay equation~\eqref{fgfam} in case $N=2$. 
It can be shown that the conjecture is true in general
for the linear delay equation 
$\dot x(t)=A x(t-\tau)$, where $x$ is a variable in $\bbR^n$ and
$A$ is a nonsingular $n\times n$ matrix~\cite{tobepub}.

As we have already discussed, singular equations
of motion like system~\eqref{vecsingsys} generally have
unphysical runaway solutions. To eliminate these
solutions and leave only the physical solutions,  the singular
system must be replaced by the 
dynamical system on the corresponding slow manifold.
In effect, the truncated equations obtained 
from the underlying delay equation after expansion
in the small delay must be replaced by 
the system obtained using iterative reduction; 
this system is equivalent to the dynamical system on the
slow manifold. Without this replacement, the
appearance of spurious runaway modes in inevitable, and 
their existence will cause overflows in numerical simulations.

\section{Gravitational radiation damping}\label{sec:3}
To illustrate the singular perturbation procedure described
in section~\ref{sec:2} as a method for
the elimination of runaway solutions, 
we examine an application of this
approach to a one-dimensional Abraham-Lorentz 
equation of the form~\eqref{5der}.

Let us consider an ideal linear quadrupole oscillator
(that is, two masses $m$ and $m'$ 
connected by a spring of negligible mass), 
where the only source of damping is the
gravitational radiation reaction force
associated with the emission of gravitational radiation
due to the variable quadrupole moment of the system.
A model for the (dimensionless) relative position $z$ of these particles,
with gravitational radiation damping included,
is the fifth-order ordinary differential equation
\begin{equation}\label{sergieos}
\mu z\frac{d^5 z^2}{dt^5}+\frac{d^2 z}{dt^2}+z=1,
\end{equation}
where the small parameter is given by 
$\mu=4 G \mu_0\ell_0^2 \omega_0^3/(15 c^5)$,
$\mu_0$ is the reduced mass ($\mu_0^{-1}=m^{-1}+m'^{-1}$),
$\ell_0$ is the spatial scale parameter and $\omega_0$ is
the frequency of the ideal linear oscillator such that
$\omega_0^{-1}$ is the temporal scale parameter.
In equation~\eqref{sergieos}, we have neglected the Newtonian 
gravitational interaction between $m$ and $m'$ as well as
all relativistic effects except for 
radiation damping. 
Let us note that this 
oscillator has an equilibrium solution $z(t)\equiv 1$.
According to our general approach described in section~\ref{sec:2},
this system can be treated as a singular perturbation
problem, where the physically correct dynamical system would be the system
defined on the slow manifold of a corresponding,
appropriately chosen, first-order system. 

We emphasize that the fifth-order differential equation~\eqref{sergieos}
is not the correct physical model;
for example, to specify a solution, the initial position and its first
four time derivatives must be given.
Even with the obvious choice for these
initial conditions---that is, the initial conditions for a sinusoidal
oscillation---a numerical integration shows that such
solutions do not oscillate; rather, they are divergent.

To obtain a system with the expected dynamical behavior of an under-damped 
oscillator, the differential equation~\eqref{sergieos} must be replaced
by its restriction to an appropriate slow manifold. We will not carry
out the complete reduction procedure here~\cite{tobepub}. 
We note, however, that 
the system matrix for the 
linearization of system~\eqref{sergieos}
at the steady state solution $z=1$
has five distinct eigenvalues that are given
to lowest order in the small parameter by
\[
-(2\mu)^{-1/3},\qquad (2\mu)^{-1/3}(\frac{1}{2}\pm i\frac{\sqrt{3}}{2}),
\qquad -\mu\pm i.
\] 
For small $\mu$, the first three eigenvalues are `fast'  and the last two
are `slow'. This suggests that the nonlinear system has a two-dimensional
slow manifold. In fact, in accordance with our general scheme,
the restriction of the dynamical system to this invariant
manifold is a second-order system that gives the correct post-Newtonian
dynamics. In this case, 
the dynamical system on the slow manifold to first order 
in the small parameter is given by the second-order 
differential equation
\begin{equation}\label{sossl}
\ddot z+32\mu z(z-\frac{15}{16})\dot z+z=1.
\end{equation} 
A unique solution of this equation is obtained by specifying only
the initial relative position and velocity of the oscillating masses.
For $z$ near the equilibrium state $z=1$, the expected dynamics for the
radiating system is revealed: the relative motion is an under-damped 
oscillator. Numerical integration of this equation using standard
algorithms is stable and produces the expected result.

The iterative reduction procedure can also be used to
obtain equation~\eqref{sossl} from equation~\eqref{sergieos}.
Even our simple example illustrates the necessity
of reducing the higher-order
equations of motion involving radiation reaction
before numerical integration. For the more realistic
hydrodynamic equations that include conservative 
post-Newtonian terms as well as 
radiation reaction, the corresponding
Euler equation must involve these forces in the \emph{reduced}
form, that is, they should contain at most the position
and the velocity of the fluid element~\cite{rezzolla, rezzolla2}. 

\section{Delay equations with sufficiently large delays}\label{sec:4}
It is important to point out that
the reduction procedure described in sections~\ref{sec:2} and~\ref{sec:3}
cannot in general be expected to produce a good approximation to the true
dynamics for `large' delays. 

As a simple but revealing example, let us reconsider the scalar 
linear delay equation $\dot x(t)=-a x(t-\tau)$ with $a>0$.
For small $\abs{\tau}$, we have already shown that all orbits in
the state space are attracted to a one-dimensional
attractor on which the dynamical
system is given by the vector field~\eqref{eq:ffe}
with $a\mapsto -a$. For $|\tau|$ less than the radius
of convergence of this series $\tau^*=(|a|e)^{-1}$, 
the correct
dynamical behavior of the delay equation is predicted by this
vector field.  Because, in this case,
the zeroth-order approximation $\dot x=-a x$ already has a hyperbolic 
structure (that is, all solutions are attracted
to the rest point at the origin exponentially fast), even the zeroth order
approximation determines the qualitative dynamics for these values of
$\tau$.
By inspection of this delay equation,
it might seem natural to conclude
that the fixed delay $\tau$ does not influence the
behavior for sufficiently large $t$ and 
the approximation $\dot x=-a x$ remains valid for all fixed delays.
This is not true.  For instance, if $\tau=\pi/(2 a)$, then
the delay equation has the two-parameter family of exact solutions
\begin{equation}\label{tpfps}
t\mapsto c_1 \cos {a t}+c_2 \sin {a t}.
\end{equation}
Therefore, the qualitative behavior of the delay equation
$\dot x(t)=-a x(t-\pi/(2a))$ is certainly not predicted by
the ordinary differential equation $\dot x=-a x$, or by the
corrections to this equation within the radius of convergence
of the slow vector field.
The transition of the dynamical behavior of this delay equation
from a stable rest point to a periodic regime as
$\tau$ increases is easily seen to be the result of
a degenerate Hopf bifurcation~\cite{mm}. 
Indeed, we recall that  $x(t)=\exp(\lambda(\tau) t)$ is
a solution of the delay equation under consideration
if $\lambda$ is a solution of the
characteristic equation $\lambda=-a \exp(-\lambda\tau)$. 
For $\tau<\tau^*$, the solutions of this equation have negative real
parts and all such solutions are therefore attracted to the zero
solution. 
If $\tau=\pi/(2 a)$, 
then the characteristic equation has a pair of
pure imaginary roots that give rise to the two-parameter family
of periodic solutions~\eqref{tpfps}. For
$\tau>\pi/(2 a)$, the characteristic equation has 
roots with positive real parts; therefore, 
there are solutions that grow without bound. 
Nevertheless, for these values of $\tau$, the delay equation
has an attractor. 
In fact, for $\pi/(2 a)\le \tau<\pi/(2a)+2\pi/a$,
there is a two-dimensional attractor and 
the dynamical system on the attractor has the
form $\ddot x-2\theta \dot x+(\theta^2+\varphi^2) x=0$ corresponding
to the roots $\lambda_\pm =\theta(\tau)\pm i \varphi(\tau)$ 
of the characteristic equation 
$\lambda=a\exp(-\lambda\tau)$
with positive real parts. 
As $\tau$ increases further, the dimension of the
attractor increases discontinuously by two at each 
$\tau=\pi/(2a)+2 N \pi/a$, where
$N=2,3,4\ldots$.
\begin{figure}[ht]
\centerline{\psfig{file=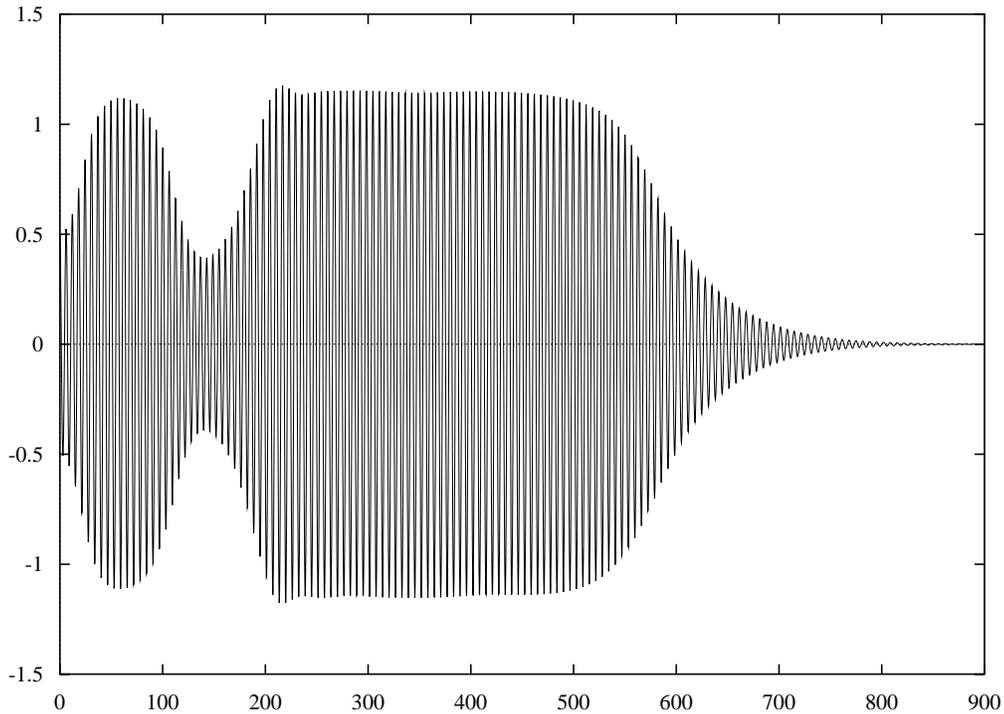, width=25pc}}
\caption{\label{fig:1} Plot of $x$ versus $t$ for the delay-differential 
equation~\eqref{dufdel}. Here 
$x(t)\equiv 0.5$ on the interval $t\le 0$, 
$w(0)=0$, $\alpha=0.1, \beta=0.1$, $\Omega=1$, $\rho=0.006$ and 
$\nu=10.5$.} 
\end{figure}

The Hopf bifurcation for delay equations with constant delays
has been studied in detail. 
For instance, a more sophisticated analysis 
(see, for example, \cite[p. 341]{hale})
shows that $\tau=\pi/2$ is a supercritical Hopf bifurcation value
for the nonlinear scalar delay equation
\[
\dot x(t)=-[1+x(t)]x(t-\tau).
\]
Moreover, this system has a nontrivial periodic orbit for each $\tau>\pi/2$. 

The delay-type equations of astrophysics generally do not have
constant delays. But as we have mentioned, 
for two coalescing neutron stars
with nearly equal masses and on nearly circular orbits, the delays involved
are almost constant;
in fact, this `fast' periodic motion evolves as a result of radiation
damping on a timescale that is much longer than $P_b$. During this
`slow' evolution,
the delay increases  as
the radius of the binary decreases due to the
emission of gravitational waves. 
Motivated by this astrophysical scenario,
we have studied an 
oscillator model with a time-dependent delay. 
This example is not intended to be
a realistic model, rather it is meant to illustrate
some of the bifurcation phenomena that occur in delay equations with
time-dependent delays. 
Our example is the second-order differential-delay equation
\begin{equation}\label{dufdel}
\nonumber \ddot x(t)+\Omega^2 x(t)+ \alpha x(t-w)-\beta x^3(t-w)=0,
\end{equation}
where $\alpha$, $\beta$ and $\Omega$ 
are constant system parameters and $x$ is viewed as
the state variable of a (Duffing) oscillator with variable delay 
$w(t)$ such that  
$\dot w(t) +\rho w(t)=\rho \nu$.
Here $\rho$ and $\nu$ are constants; hence, 
$w(t)-\nu$ is an exponentially decreasing or increasing
function of time depending on whether  $\rho$ is positive
or negative, respectively. In any case, we have
a dynamic delay that is asymptotic to the constant value $\nu$. 
Note that if $\rho=0$, the delay is constant; in this case,
the corresponding second-order differential equation on the
slow manifold (to first order in $w$) is given by
\begin{equation}\label{vander}
\ddot x+ w (3 \beta x^2-\alpha)\dot x+(\alpha+\Omega^2)x -\beta x^3=0,
\end{equation}
a form of van der Pol's equation. In case $\Omega\ne 0$, this
differential equation typically has a stable
limit cycle for $w>0$. But for $\Omega=0$ 
(that is when all forces are retarded),
it is easy to prove that no periodic orbits exist and most solutions
are unbounded.

A typical plot of $x$ versus $t$ for 
system~\eqref{dufdel} for $\Omega^2-\alpha>0$ is given in 
figure~\ref{fig:1}, where
the delay increases from its initial value $w(0)=0$ to $\nu=10.5$.
The initial response of the system (where the delay is small)
is characterized by an oscillation as expected from
equation~\eqref{vander}, which follows from the expansion
of equation~\eqref{dufdel} to first order in $w\ll 1$.
But as $w$ increases, the qualitative behavior of the system is
affected by three additional bifurcations not accounted for
by equation~\eqref{vander}. At the third bifurcation,
the stable oscillation disappears. 
Additional bifurcations of the same type occur if $\nu$ is set
to a larger value. 
Numerical experiments suggest that these 
bifurcations are not Hopf bifurcations;
instead, they are `center bifurcations', where at some
parameter value there is a rest point of center
type and one of the periodic orbits surrounding this rest point 
continues to exist as the parameter is changed. 
The family~\eqref{vander} with
the parameter values as in figure~\ref{fig:1} has
a bifurcation of this type as $w$ increases through
$w=0$.

The behavior depicted in figure~\ref{fig:1} is suggested by an
analysis of the roots of the characteristic equation, 
\[ \zeta^2+\Omega^2+\alpha e^{-\zeta w}=0,\]
for the linearization of the delay equation~\eqref{dufdel}.
The bifurcation points (corresponding to the existence of centers) 
are given by
\begin{equation}\label{bifpoints}
w_\ell=\ell \pi (\Omega^2+\alpha \cos \ell \pi)^{-1/2},
\end{equation}
where $\ell$ is a non-negative integer. These are the values of $w$
such that the characteristic equation has pure imaginary roots.
A computation shows that if $\ell$ is even, then as $w$ increases 
a pair of pure imaginary roots crosses the imaginary axis into the
right half-plane, and  if $\ell$ is odd, then the roots cross into
the left half-plane. Under the assumption that the 
bifurcations are supercritical, a stable limit cycle appears
after the bifurcation in the first case; in the second case, 
a stable limit cycle disappears. 
For the parameter values used to obtain figure~\ref{fig:1},
the bifurcation values computed from equation~\eqref{bifpoints}
are (approximately) $0$, $3.3$, $6.0$, $9.9$ for $\ell=0,1,2,3$
such that $w_\ell\le \nu$.
At $w_0=0$ a limit cycle appears, at $w_1\approx 3.3$ the limit cycle
disappears, and so on. Thus, these bifurcations account for the
appearance and disappearance of oscillations in figure~\ref{fig:1}.
We note that a similar sequence of bifurcations occurs
whenever $\Omega^2-\alpha>0$. On the other hand,
if $\Omega^2-\alpha<0$ (for example if $\Omega=0$), 
then all bifurcation points correspond to roots of the
characteristic equation crossing into the right half-plane.
In this case, the bifurcations can be subcritical. Indeed,
for $\Omega=0$, numerical simulations indicate that no limit cycle
appears. As a result, solutions starting near the unstable rest point
become unbounded.

The slow dynamical system, obtained by reduction from a truncation of
an expansion of a delay equation in powers of the delay,
approximates the dynamics on the global attractor of the delay equation
as long as the delay is sufficiently small; 
but, as our examples show,
the ordinary differential equations obtained by expansion,
truncation, and reduction \emph{cannot} be used in general to predict 
the correct dynamical behavior for sufficiently large delays. 
We have mentioned, for example, that the 
dimension of the attractor of a family of delay equations, parametrized
by the delay, can increase in dimension so that 
that the corresponding slow vector field is no longer defined
on the attractor.
But this is not the only possible
scenario for the appearance of new attractor dynamics; for example, 
the attractor could cease to exist or be a manifold
for some values of the delay.

The Abraham-Lorentz type equation~\eqref{5der} can be used,
after reduction to a slow manifold, 
to predict the relative orbital motion of a 
relativistic binary system in the regime where 
the delay is sufficiently small. 
The size of the maximum allowed delay would have
to be computed on a case-by-case basis using the explicit
form of the delay equation that models the dynamics of
a coalescing pair of neutron stars.
The results of this section show that for
sufficiently large delay the attractor does not in general correspond to
the slow manifold. The question remains whether such a divergence of
behaviors could ever occur in the case of retarded equations of classical
field theory. This is an interesting open problem.

\section{Discussion} \label{sec:5}
It is expected that interferometric gravitational
wave detectors that are presently under construction
will be able to detect signals from massive coalescing binary systems. 
For the analysis of such forthcoming data,
it is important to have theoretically predicted wave forms
(`templates') for the relevant astrophysical processes.
To this end, extensive computations are necessary that need to take
gravitational radiation reaction into 
account~\cite{rezzolla,rezzolla2, rezzolla3, rezzolla4}.
The standard approach leads to higher time-derivative equations
that involve runaway modes and inevitably produce incorrect results.

We have determined the source of the difficulty 
by investigating delay equations, which are essentially nonlocal,
and the higher time-derivative equations that are obtained by
truncations of the
(post-Newtonian) expansions in powers of the delay.
For sufficiently small delays, a proper justification is provided 
for the usual method of replacing terms with higher-derivatives by terms 
with at most first derivatives using
repeated substitution of the equations of motion
(`iterative reduction'). We have shown that
in the investigation of the solutions of 
higher-derivative equations that represent phenomena
involving radiation reaction, it is essential
to reduce such equations to the corresponding slow manifolds before
numerical analysis. Our work suggests, however, that unexpected
nonlocal phenomena could occur for sufficiently large delays
that cannot be predicted using the
local equations of motion even after iterative reduction.

\end{document}